\documentclass[12pt,preprintnumbers,eqsecnum]{revtex4}
\usepackage{amsfonts}
\usepackage{mathrsfs}
\usepackage{amssymb}
\usepackage{amsmath}
\usepackage{graphicx,epsfig}
\usepackage{graphicx}
\usepackage{epstopdf}
\usepackage{pstricks}
\usepackage{slashed}
\usepackage{mathbbol}
\usepackage{color}
\usepackage{slashed}
\def\be{\begin{equation}}
\def\ee{\end{equation}}
\def\bea{\begin{eqnarray}}
\def\eea{\end{eqnarray}}
\def\bean{\begin{eqnarray*}}
\def\eean{\end{eqnarray*}}
\def\bary{\begin{array}}
\def\eary{\end{array}}
\def\bit{\begin{itemize}}
\def\eit{\end{itemize}}

\def\su5u1{SU(5) \times U(1)}
\def\fsu5u1{SU(5) \times U(1)'}
\def\so10{SO(10)}
\def\sq20{SO(10) \times SO(10)}

\def\bwt{\begin{widetext}}
\def\ewt{\end{widetext}}
\def\be{\begin{equation}}
\def\ee{\end{equation}}
\def\bea{\begin{eqnarray}}
\def\eea{\end{eqnarray}}
\def\bean{\begin{eqnarray*}}
\def\eean{\end{eqnarray*}}
\def\bary{\begin{array}}
\def\eary{\end{array}}
\def\bit{\begin{itemize}}
\def\eit{\end{itemize}}

\def\su5u1{SU(5) \times U(1)}
\def\fsu5u1{SU(5) \times U(1)'}
\def\so10{SO(10)}
\def\sq20{SO(10) \times SO(10)}

\usepackage{amsmath}

\usepackage{amssymb}
\usepackage{gensymb}

\begin{document}
\setlength{\parskip}{0.1cm}

\preprint{ OSU-HEP-15-05}

\title{\large Superworld without Supersymmetry}

\author{Shreyashi Chakdar}
\email{chakdar@okstate.edu}

\author{Kirtiman Ghosh}
\email{kirti.gh@gmail.com}

\author{S. Nandi}
\email{s.nandi@okstate.edu}
\affiliation{Department of Physics and Oklahoma Center for High
Energy Physics, Oklahoma State University, Stillwater, OK
74078-3072, USA}
%\date{\today}
%%%%%%%%%%%%%%%%%%%%%%%%%%%%%%%%%%%%%%%%%%%%%%%%%%%%%%%%%%%%%%%%%%%%%%%%%%%%
\begin{abstract}
%%%

It is a possibility that the superworld (supersymmetric partners of our world) does exist without supersymmetry. The two worlds are being distinguished by an unbroken discrete $Z_2$ symmetry (similar to R-parity in supersymmetry). We lose the solution to the hierarchy problem. However, such a scenario has several motivations. For example, the lightest neutral superworld particle will be a candidate for dark matter. The other being, as in supersymmetry, it is possible to achieve gauge coupling unification. One major difference with the supersymmetric theory is that such a theory is much more general since it is not constrained by supersymmetry. For example, some of the gauge couplings connecting the Standard Model particles with the superpartners now become free Yukawa couplings. As a result, the final state signals as well as the limits on the superworld particles can be modified both qualitatively and quantitatively. The reach for these superworld particles at the Large Hadron Collider (LHC) can be much higher than the superpartners, leading to the increased possibility of discovering new physics at the LHC.

\end{abstract}



\maketitle

%%%%%%%%%%%%%%
%\newpage
%%%%%%%%%%%%%%

\section{Introduction}

The Standard model (SM) has been extremely successful to explain essentially all the experimental observations so far upto few hundred GeV scale. However SM can not accommodate the non-zero neutrino masses and the existence of the dark matter. There are also several theoretical drawbacks of the SM, such as the charge quantization, strong CP problem, baryon asymmetry and hierarchy problem. Supersymmetry (SUSY) (see e.g., \cite{Martin:1997ns},\cite{Haber:1984rc}) is a very elegant extension of the SM to solve the  hierarchy problem, as well as giving a natural dark matter candidate. SUSY requires two SM Higgs doublets, as well as bosonic (fermionic) superpartners for each SM particles. Intense experimental effort  at LEP2 and the  Tevatron was carried out to detect signals of supersymmetry. Now  at much higher energy machine at the Large Hadron Collider (LHC),  such intense effort is being continued  to discover these superpartners, but so far with no success. Several limits have been set  on the masses of these superpartners, such as gluinos, squarks, and charginos, the limits depending somewhat on the SUSY scenario being considered. For example, current limit on the gluino mass in the case of squarks being very heavy is about $1.3$ TeV \cite{a7875,a05525,a03555,c019}; while the limit on the squark mass in the limit of very heavy gluino mass is about $850$ GeV \cite{a7875,c019}, while for the equal mass gluinos and squarks, the limit is $1.8$ \cite{a05525}. The limits on the electroweak superpartners are much lower, because of their low production cross sections. For example, the limit on the lighter chargino from the 8 TeV LHC is only about $400$ GeV \cite{a5294,a7029,c006}.
In this work, we propose a model in which superpartner particles do exist in their own world and we call it to be superworld. Our world is the observed SM particles (with two Higgs doublets as in minimal supersymmetric SM (MSSM)). The two worlds are distinguished by a discrete unbroken $Z_2$ symmetry (similar to R-parity in MSSM). Because we give up supersymmetry, we  assume that observed  EW scale Higgs mass \cite{Aad:2015zhl} is fine tuned. However, the model still has several motivations. One is that because of the exact $Z_2$ symmetry, the lightest neutral particle in the superworld is a candidate for the dark matter. The gauge coupling unification can be achieved by choosing suitably the superworld particle spectrum and the couplings (which were gauge couplings because of supersymmetry), but now become Yukawa couplings. This theory/model is much less restrictive compared to supersymmetric theory. The reason being that the mass spectrum is not constrained by supersymmetry, some of the couplings between the SM particles and the superpartners, which were equal to the gauge couplings because of the supersymmetry, now become undetermined Yukawa couplings.  As a result, these can be much larger than the gauge couplings leading to much bigger cross section for their productions at the LHC. Thus the mass reach for these particles at the LHC will be much higher. Also  the missing energy of the events, as well as the missing $p_T$'s of the final state jets will be much harder than those expected in the case of supersymmetry.

\section{The model and the formalism}

Our gauge symmetry is $SU(3)_C  \times  SU(2)_L  \times U(1)_Y   \times  Z_2$. The particle content of the model is the same as minimal supersymmetric Standard Model (MSSM) broken into two worlds.
Our world is usual fermions, gauge bosons, and two Higgs doublets. The superworld contains these supersymmetric partners of our world particles. But the two worlds are not connected by supersymmetry.  The particles, their representations under the gauge group, and their $Z_2$  quantum numbers are listed in Table (\ref{Table1}).

\begin{table}[htb]
%\hskip -10pt
\begin{center}
\begin{tabular}{|c|c|c|}
\hline 
&Our World($Z_2=+1$) & Superworld ($Z_2=-1$)\\\hline
\small Matter &${\begin{pmatrix} u \\ d \end{pmatrix}}_L\sim(3,2,\frac{1}{6}), u_R\sim (3,1,\frac{2}{3}), d_R \sim (3,1,-\frac{1}{3})$ & \small{${\begin{pmatrix} \tilde{u} \\ \tilde{d} \end{pmatrix}}_L\sim (3,2,\frac{1}{6}), \tilde{u_R}\sim (3,1,\frac{2}{3}), \tilde{d_R}\sim (3,1,-\frac{1}{3})$}\\
&$ {\begin{pmatrix} \nu_e \\ e \end{pmatrix}}_L\sim (1,2,-\frac{1}{2}), e_R\sim (1,1,-1), \nu_R\sim (1,1,-1)$ & ${\begin{pmatrix} \tilde{\nu_e} \\ \tilde{e} \end{pmatrix}}_L\sim (1,2,-\frac{1}{2}), \tilde{e_R}\sim (1,2,-1), \tilde{\nu_R}\sim (1,2,-1)$\\
%&$ {\begin{pmatrix} \nu_\tau \\ \tau \end{pmatrix}}_L, \tau_R$ & ${\begin{pmatrix} \tilde{t} \\ \tilde{b} %\end{pmatrix}}_L, \tilde{b_R}, \tilde{t_R}, {\begin{pmatrix} \tilde{\nu_\tau} \\ \tilde{\tau} \end{pmatrix}}_L, %\tilde{\tau_R}$\\
\hline
\small Gauge & $G_{a,a=1-8}, A_{i, i=1-3}, B$ & $\tilde{g}_{a,a=1-8}, \tilde{A}_{i, i=1-3}, \tilde{B}$\\
\hline
\small Higgs & $H_u$, $H_d$ & $\tilde{H}_u$, $\tilde{H}_d$\\
\hline
\end{tabular}
\end{center}
\caption{Matter, gauge and higgs contents of our world and the superworld.}
\label{Table1}
\end{table}

All the particles in our world  have $Z_2 = +1$, whereas all the particles in the superworld has $Z_2 = - 1$. Note that we need two Higgs doublets as in MSSM to cancel the gauge anomaly among the Higgssinos.  We have two Higgs doublets, $H_u$ and $H_d$.   We assume that only one Higgs doublet $H_u$ couples to the fermions. To achieve that, we impose a discrete symmetry $Z'_2$ under which $H_d \rightarrow -H_d$, whereas all other particles of our world and superworld remains uncharged. This will avoid unwanted flavor changing neutral current at the tree level.  We note that the Higgs potential is general here, and the quadratic terms in the Higgs potential are not coming from the D-terms as in supersymmetry. The Lagrangian for our model contains three parts. The Lagrangian for our world, $L_O$, for the superworls, $L_S$, and the interaction between the two worlds, $L_I$.
The Lagrangian for our world $L_o$ is the same as is the Standard Model except for the Higgs potential. The Higgs potential for the two doublets $H_u$ and $H_d$ is given by,
\begin{eqnarray}
V=&-&\left( \mu_u^2 H_u^\dagger H_u + \mu_d^2 H_d^\dagger H_d\right ) - \mu_{12} \left(H_u^T H_d + h.c \right) + \lambda_1\left(H_u^\dagger H_u \right)^2 + \lambda_2\left(H_d^\dagger H_d \right)^2 \nonumber\\ &+& \lambda_3 \left(\left(H_u^T H_d \right)^2 + h.c \right)+\lambda_4 \left(\left(H_d^T H_u \right)^2 + h.c \right) + \lambda_5 \left(\left(H_u^T H_d\right)\left(H_d^T H_u\right) + h.c \right)
\end{eqnarray}
Note that this is somewhat different than the usual two higgs doublet model with the same symmetry of $H_d \rightarrow -H_d$. This is because $H_u$ has hypercharge Y=+1, whereas $H_d$ has hypercharge Y=-1, so that the corresponding Higgsinos cancel the gauge anomalies in the superworld.\\
The Yukawa interactions for the ordinary world is same as in the Standard Model and is given by
\begin{eqnarray}
{\cal L_o}^{\text{yuk}} &\supset&  \bar{q}_L Y_d H_u d_R +  \bar{q}_L Y_u u_R i \sigma_2 H_u^*  + \bar l_L Y_l e_R H_d + {\rm h.c.}
\end{eqnarray} 
where $Y_u$ and $Y_d$  and $Y_l$ are up type quark , down type quark and  charged lepton Yukawa coupling matrices.\\

The matter kinetic terms for the superworld is given by,
\begin{eqnarray}
{\cal L}_{\text{matter}}^{\text{kin}} \supset &+& \left({\cal D}_\mu\bar{q}_{Li}\right)^\dagger\left({\cal D}^\mu q_{Li}\right)+ \left({\cal D}_\mu\tilde{d}_{Ri}\right)^\dagger\left({\cal D}^\mu \tilde{d}_{Ri}\right)
+ \left({\cal D}_\mu\tilde{u}_{Ri}\right)^\dagger\left({\cal D}^\mu \tilde{u}_{Ri}\right)\nonumber\\
&+& \left({\cal D}_\mu\tilde{l}_{Li}\right)^\dagger\left({\cal D}^\mu \tilde{l}_{Li}\right)
+\left({\cal D}_\mu\tilde{e}_{Ri}\right)^\dagger\left({\cal D}^\mu \tilde{e}_{Ri}\right) \nonumber\\
&-& \tilde m_{q_L}^2\tilde q_L^\dagger \tilde q_L - \tilde m_{u_R}^2\tilde u_R^\dagger \tilde u_R - \tilde m_{d_R}^2\tilde d_R^\dagger \tilde d_R - \tilde m_{l_L}^2\tilde l_L^\dagger \tilde l_L - \tilde m_{e_R}^2\tilde e_R^\dagger \tilde e_R\label{ktl}
\label{matter}
\end{eqnarray}
where $\cal D_{\mu}$ is the usual gauge covariant derivative and i = 1,2,3 represent three families.

\begin{eqnarray}
{\cal L}_{\text{gauge}}^{\text{kin}} \supset &+& \frac{1}{2} i \tilde{G}^T_a \left( \slashed{D} \tilde{G} \right)_a  - \frac{1}{2} \text{m}_{\tilde{G}} \tilde{G}^T_a C^{-1}\tilde{G}_a 
+ \frac{1}{2} i \tilde{A}^T_i \left( \slashed{D} \tilde{A} \right)_i  - \frac{1}{2} \text{m}_{\tilde{A}} \tilde{A}^T_i C^{-1}\tilde{A}_i \nonumber\\
&+& \frac{1}{2} i \tilde{B}^T \left( \slashed{D} \tilde{B} \right)  - \frac{1}{2} \text{m}_{\tilde{B}} \tilde{B}^T C^{-1}\tilde{B} + i \bar{\tilde{H_u}}\slashed{D} \tilde{H_u} + i \bar{\tilde{H_d}}\slashed{D} \tilde{H_d}
+ \mu \tilde{H_u}^T i \sigma_2 \tilde{H_d}
\label{gauge}
\end{eqnarray}

where $\cal D_{\mu}$ = $\gamma^\mu D_\mu$ is the usual covariant derivative with the indices a = 1-8 for eight gluinos and i = 1-3 for three $SU(2)_L$ gauginos. Note that as in supersymmetry, we have assumed the gauginos to be Majorana particles.\\
The gauge interactions between our world and the superworld is contained in Eqns (\ref{matter}) and (\ref{gauge}) coming from the covariant derivatives. In addition, we have Yukawa interactions between our world and the superworld. These interactions are given by,
\begin{eqnarray}
{\cal L_{\text{O-S}}}^{\text{yuk}} &\supset& \lambda_{\tilde{q}} \bar{q}_L \tilde{G} \tilde{q}_L +\lambda_{\tilde{u}} \bar{u}_R \tilde{G} \tilde{u}_R + \lambda_{\tilde{d}} \bar{d}_R \tilde{G} \tilde{d}_R  + {\rm h.c.}
\label{5}
\end{eqnarray} 
 In Eq.~\ref{5}, for simplicity we have written only the first family couplings.\\
Note that all the couplings in Eq.~\ref{5}  are gauge couplings in supersymmetric theory. In our theory, they are independent gauge couplings and can be much larger. These will give rise to much bigger cross-sections for producing superworld particles at the colliders.
Finally, for the superworld particles, we have usual mass terms which are independent parameters and are not constrained by supersymmetry.

\section{Phenomenology}

In this section, we will discuss the phenomenology of this model in the context of the LHC experiment. The particle spectrum of our model is similar to the particle spectrum of supersymmetric scenarios. However, we have not assumed any supersymmetry like symmetry connecting  our world and the superworld. Therefore, apart from the gauge interactions, all other interactions between our world and superworld are not constrained by any symmetry. As a result, this model could potentially gives rise to collider signatures which are very different from supersymmetric scenarios. Before going in to the discussion of collider signatures, it is important to understand how the production cross-sections of super-particles at the LHC are modified in absence of supersymmetry between our world and the superworld. 

%------------------------------------------------------------------- 
\begin{figure}[t]
\begin{center}
\epsfig{file=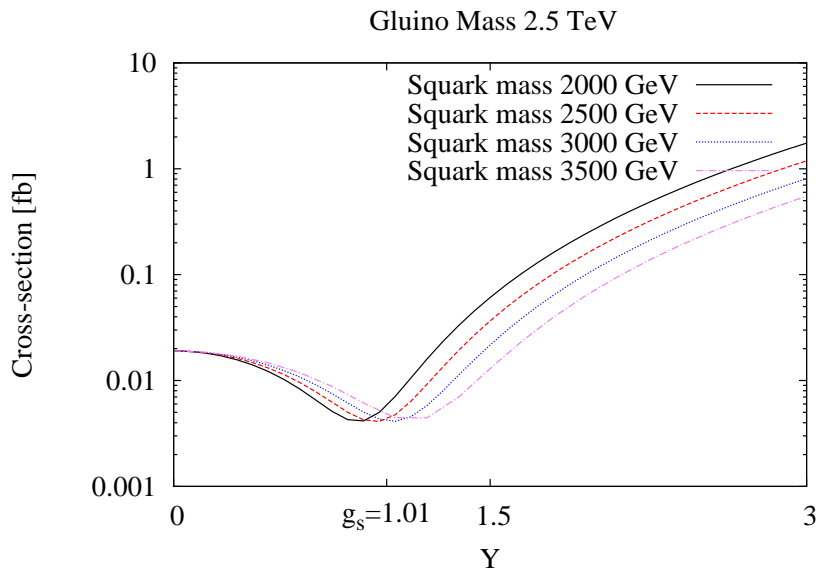,width=8cm,angle=270}
\end{center}
\caption{Gluino pair production cross-section as a function of gluino-squark-quark Yukawa coupling ($Y$) for four different values of squark mass. Gluino mass is assumed to be 2.5 TeV. We have used CTEQ6L1 \cite{Nadolsky:2008zw} parton distribution functions with the factorization scale (for parton distribution functions) and scale of $\alpha_s$ fixed at the parton center-of-mass energy ($\sqrt {\hat s}$). On the x-axis, we have indicated the point corresponds to $Y=g_S({\rm at~the~gluino~mass})$. The cross-sections corresponding to this point represent gluino pair production cross-sections in supersymmetric scenarios.}
\label{gluino}
\end{figure}
%-------------------------------------------------------------------

The LHC is a proton-proton collider. Therefore, the strongly interacting particles, in this case, squarks ($\tilde q$) and gluinos ($\tilde g$), are copiously prodused at the LHC. In our model, we have demanded a $Z_2$ symmetry with $Z_2=+1$ for all particles in our world and $Z_2=-1$ for all particles in the superworld. The phenomenological consequences of this $Z_2$ symmetry is similar to the consequences of $R$-parity in supersymmentric scenarios. For example, $Z_2$ symmetry allows the decay of a super particle only into a lighter super particle in association with one or more SM particles. This makes the lightest particle of superworld stable and hence, a good candidate  for the  cold dark matter. Moreover, as a consequence of this $Z_2$ symmetry, super particles can only be pair produced. In this work, we have studied the pair production of superparticles in the framework of our  model which does not assume supersymmetry. We have estimated the pair production cross-sections at the LHC with 13 TeV center-of-mass energy. We have used a parton-level Monte-Carlo code to numerically integrate over the phase-space and parton distributions. CTEQ6L1 \cite{Nadolsky:2008zw} parton distribution functions are used with the factorization scale (for parton distribution functions) and scale of $\alpha_s$ fixed at the parton center-of-mass energy ($\sqrt {\hat s}$).

%------------------------------------------------------------------- 
\begin{figure}[t]
\begin{center}
\epsfig{file=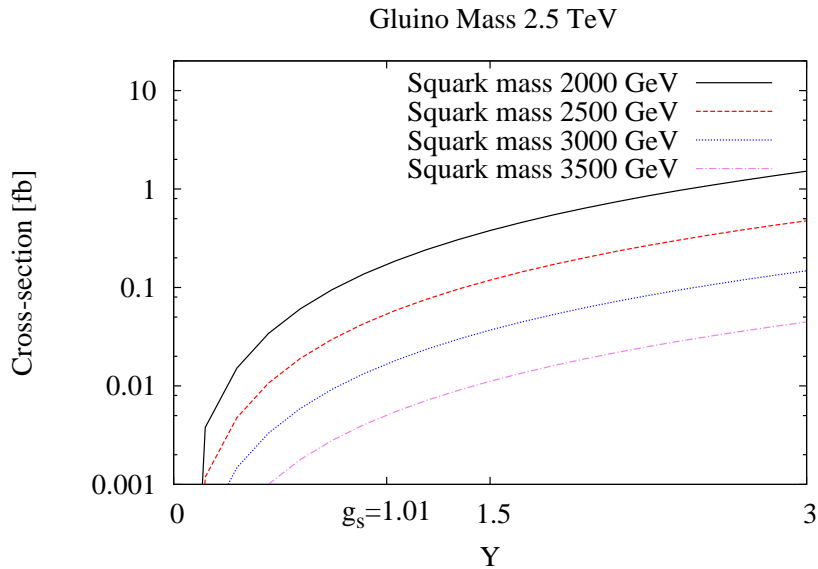,width=8cm,angle=270}
\end{center}
\caption{Same as Fig.~\ref{gluino} but for Squark-gluino pair production for a 2.5 TeV gluino and for four different values of squark mass. }
\label{qg}
\end{figure}
%-------------------------------------------------------------------

\begin{itemize}
\item Let us begin our discussion with gluino pair production. A pair of gluino can couple to a gluon. In the framework of this model, this is a gauge coupling arises from the kinetic term of gluinos. Gluinos also have tree level Yukawa couplings with a squark and a SM quark. Unlike supersymmetry, the strengths of these Yukawa couplings are not constrained to be equal to  the gauge couplings. At the LHC, contributions to the gluino pair production come from gluon-gluon and quark anti-quark initiated processes. At the parton level, $gg \to \tilde g \tilde g$ process is mediated by 
%gauge interactions only with 
a gluon in the $s$-clannel or a gluino in the $t$-channel; whereas, the Feynman diagrams $q \bar q \to \tilde g \tilde g$ include both gauge interactions (for $s$-clannel gluon exchange) and Yukawa interactions ($t~(u)$-channel squark exchange). These Yukawa couplings are free parameters in our model.  In Fig.~\ref{gluino}, we have presented gluino pair production cross sections for a 2.5 TeV gluino as a function of gluino-squark-quark Yukawa coupling ($Y$) for different values of squark mass. Fig.~\ref{gluino} shows a minima which can be attributed to the fact that gauge interaction mediated diagram ($s$-channel gluon exchange diagram) and Yukawa interaction mediated diagrams ($t-(u)$-channel squark exchange) in $q \bar q \to \tilde g \tilde g$ process interefare destructively. The minima in Fig.~\ref{gluino} corresponds to the particular value of the Yukawa coupling for which the destructive interference between $s$-clannel and $t-(u)$-channel diagrams reaches the maximum. Fig.~\ref{gluino} also shows that depending on the value of the Yukawa coupling, our model could give rise to a gluino pair production cross-section which is orders of magnitude larger than the gluino production cross sections in supersymmetric scenarios. For example, as can be seen from Fig.~\ref{gluino},  for a gluino mass of $2.5$ TeV, when the Yukawa coupling is equal to the gauge coupling, $g_s = 1.014$, the values of these cross sections are of the orders of $10^{-2}$ fb, while in our model, for $Y = 3$, the cross sections are as large as 1 fb.

%------------------------------------------------------------------- 
\begin{figure}[t]
\begin{center}
\epsfig{file=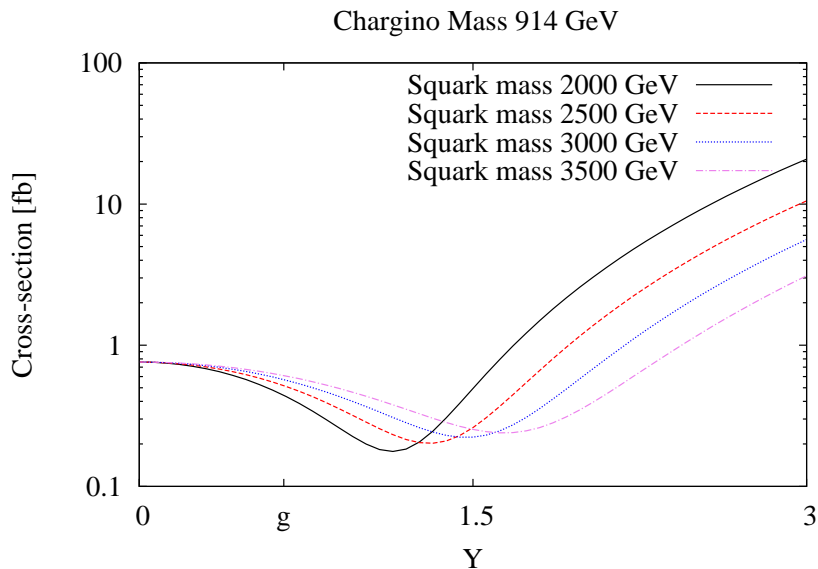,width=8cm,angle=270}
\end{center}
\caption{Chargino pair production cross-section for a 914 GeV chargino as a function of chargino-squark-quark Yukawa coupling ($Y$) for four different values of squark mass.}
\label{chargino}
\end{figure}
%-------------------------------------------------------------------

\item In Fig.~\ref{qg}, we have presented squark-gluino pair production cross-section as a function of the Yukawa coupling between gluino, squark and a quark. Fig.~\ref{qg} shows that the cross-section increases with the increasing value of the Yukawa coupling. At the parton level, $gg \to \tilde g \tilde g$ process is mediated by a $s$-channel quark exchange and $u~(t)$-channel gluino (squark) exchange. It is important to note that both Feynman diagrams contain one vertex proportional the gauge coupling and one vertex proportional to the Yukawa coupling. Therefore, the cross-section, in this case, is directly proportional to the Yukawa coupling square which explains the nature of Fig.~\ref{qg}.  Again we see that cross sections can be more than an order of magnitude larger than the supersymmetric case for large values of the Yukawa couplings.

\item In the electroweak sector of the superworld, we have studied chargino pair production. The results are presented in Fig.~\ref{chargino}. We have assumed the chargino mass to be equal to 914 GeV which results from the diagonalizitation of the chargino mass matrix. In the framework of this model, the structure of the chargino mass matrix is slightly different from the supersymmetry. In the $2\times 2$ chargino mass matrix, the diagonal elements are  given by the $SU(2)_L$ gaugino mass, $m_{\tilde A}$ (see Eq. \ref{gauge}) and $\mu$ which has analogs in supersymmetric scenarios. However, the off diagonal terms in our model are proportional to Yukawa couplings between higssino, wino and Higgs. In our analysis, we have assumed $m_{\tilde A}=\mu=1$ TeV and the Yukawa couplings in the off-diagonal terms are equal to the $SU(2)_L$ gauge coupling $g$. With these set of parameters, the diagonalization of chargino mass matrix gives rise to a light chargino ($\tilde \chi_1^{\pm}$) with mass 914 GeV. Chargino pair production at the LHC is a quark anti-quark initiated process. At the parton level, $q\bar q \to \tilde \chi_1^+ \tilde \chi_1^-$ process is mediated by $s$-channel $\gamma$ or $Z$-boson exchange and $t$-channel squark exchange. The $s$-channel and  $t$-channel amplitudes interfere destructively and hence, we observe a minima in charrgino pair-production cross-sections in Fig.~\ref{chargino}. Again, as can be seen, the chargino pair production cross sections can be orders of magnitude larger than for the supersymmetric case.

\end{itemize}  

From the above discussion, it is clear that the production cross section of a pair of superworld particles, in our model,  could be significantly different compared to a supersymmetric scenario. However, in a collider experiment, we  observe only the decay products of superparticles. Therefore, it is important to discuss the decays of superparticles in our model. In the presence of the unbroken $Z_2$-symmetry, the decay patterns of the superparticles in our model are very similar to the decay patterns predicted in different supersymmetric scenarios with conserved $R$-parity. The decays of the super particles crucially depend on the mass hierarchy in the superworld. For example, in a scenario with squarks being lighter than gluinos, gluinos dominantly decays into a quark-squark pairs and squarks subsequently decays into a chargino or neutralino in association with a SM quark. Whereas, in a scenario with gluinos being lighter than the squarks, squarks dominantly decay into a quark-gluino pair followed by a subsequent tree level three body decay of the gluinos into a chargino or neutralino in association with a fermion anti-fermion pair. It is important to note that all the decays discussed above are mediated by Yukawa interactions in our model. The Yukawa couplings are free parameters in this model. Therefore, our model could potentially gives rise to a flavor biased signature at the collider experiments which is not the case in supersymmetry.

\section{Conclusions}

We have presented a model in which the superpartners of the observed particles (we call those to be superparticles, and the collection of those particles to be the superworld) do exist, but they are not connected by supersymmetry. Thus the interactions of these superworld  particles with the observed particles (our world) is not restricted by supersymmetry and hence are much more general. In such a model, although we lose the solution to the hierarchy problem, we still have the cold dark matter, and the gauge coupling unification can be achieved with suitable choice of parameters (namely the Yukawa couplings and the masses of the superparticles). Our model is particularly interesting from the new physics discovery point of view. At the LHC, the cross sections for producing these superparticles can be orders of magnitude larger than in supersymmetry, and thus extend greatly their reach. The subsequent decays of these superparticles will give rise to similar final states as in supersymmetry, namely multijets or multileptons with high $p_T$ plus large missing energy $\slashed{E}_T$. However, because of the larger cross sections, we can produce much heavier particles, the $p_T$'s and the missing energies for the final state events will be much larger from their subsequent decays. Thus, if such signals for new physics are seen and can not be explained by supersymmetry, our model will be a good candidate. Also,because of  much larger mass reach, the discovery potential of the superparticles, and hence the new physics will be greatly enhanced at the LHC.

\section {Acknowledgement}
We thank K. S. Babu, J. Haley and R. N. Mohapatra for useful discussions. This work is supported in parte by the US Department of Energy Grant Number DE-SC0010108.

%\{Bibliography}

\end{document}